\documentclass[twocolumn,amsmath,amssymb,floatfix,shownopacs,footinbib,superscriptaddress]{revtex4-1}
\usepackage{bm,graphicx,url,epsf,color}
\usepackage{wasysym}
\usepackage{MnSymbol}

\begin{document}

\title{Transferring the quantum state of electrons across a Fermi sea with Coulomb interaction}

\author{H. Duprez}
\email{These authors contributed equally to this work.}
\affiliation{Centre de Nanosciences et de Nanotechnologies (C2N), CNRS, Univ Paris Sud, Universit\'e Paris-Saclay, 91120 Palaiseau, France}
\author{E. Sivre}
\email{These authors contributed equally to this work.}
\affiliation{Centre de Nanosciences et de Nanotechnologies (C2N), CNRS, Univ Paris Sud, Universit\'e Paris-Saclay, 91120 Palaiseau, France}
\author{A. Anthore}
\affiliation{Centre de Nanosciences et de Nanotechnologies (C2N), CNRS, Univ Paris Sud, Universit\'e Paris-Saclay, 91120 Palaiseau, France}
\affiliation{Univ Paris Diderot, Sorbonne Paris Cit\'e, 75013 Paris, France}
\author{A. Aassime}
\affiliation{Centre de Nanosciences et de Nanotechnologies (C2N), CNRS, Univ Paris Sud, Universit\'e Paris-Saclay, 91120 Palaiseau, France}
\author{A. Cavanna}
\affiliation{Centre de Nanosciences et de Nanotechnologies (C2N), CNRS, Univ Paris Sud, Universit\'e Paris-Saclay, 91120 Palaiseau, France}
\author{U. Gennser}
\affiliation{Centre de Nanosciences et de Nanotechnologies (C2N), CNRS, Univ Paris Sud, Universit\'e Paris-Saclay, 91120 Palaiseau, France}
\author{F. Pierre\thanks{frederic.pierre@c2n.upsaclay.fr}}
\email[e-mail: ]{frederic.pierre@c2n.upsaclay.fr}
\affiliation{Centre de Nanosciences et de Nanotechnologies (C2N), CNRS, Univ Paris Sud, Universit\'e Paris-Saclay, 91120 Palaiseau, France}

\begin{abstract}
The Coulomb interaction generally limits the quantum propagation of electrons. 
However, it can also provide a mechanism to transfer their quantum state over larger distances.
Here, we demonstrate such a form of teleportation, across a metallic island within which the electrons are trapped much longer than their quantum lifetime.
This effect originates from the low temperature freezing of the island's charge $Q$ which, in the presence of a single connected electronic channel, enforces a one-to-one correspondence between incoming and outgoing electrons.
Such high-fidelity quantum state imprinting is established between well-separated injection and emission locations, through two-path interferences in the integer quantum Hall regime. 
The added electron quantum phase of $2\pi Q/e$ can allow for strong and decoherence-free entanglement of propagating electrons, and notably of flying qubits.
\end{abstract}

\maketitle

A disordered environment, with a large number of interacting degrees of freedom, is generally considered as the nemesis of quantum technologies.
This is exemplified by a metallic island, often pictured as a reservoir of thermal electrons, with its large energy density of states $1/\delta$ and limited number $N$ of connected electronic channels.
Indeed, the interval between inelastic collisions destroying the quantum coherence of the electrons \cite{Nazarov2009,Pierre2003} is typically much smaller than their dwell time inside the island ($\tau_\mathrm{D}=h/N\delta$ for perfect channels \cite{Brouwer1997}, with $h$ the Planck constant).
However, we show experimentally that the Coulomb interaction in such an island can, under the right circumstances, lead to a near perfect preservation of the quantum state of electrons transferred across it.
In the employed quantum Hall regime implementation, where injection and emission points are physically separated by chirality, this constitutes a form of teleportation of the electrons' states without transmitting the physical particles themselves.
This phenomenon is different from the standard `quantum teleportation' protocol \cite{Bennett1993}, and similar to the `electron teleportation' proposed in \cite{Fu2010}.

The voltage probe model of a metallic Fermi sea \cite{Buttiker1988a} is widely used to mimic the electrons' quantum decoherence and energy relaxation toward equilibrium (see e.g. \cite{Blanter2000} and references therein).
However, independent absorption and emission of electrons result in fluctuations of the total island charge $Q$, with a characteristic charging energy $E_\mathrm{C}=e^2/2C$ (with $C$ the geometrical capacitance of the island and $e$ the elementary electron charge).
At low temperatures $T\ll E_\mathrm{C}/k_\mathrm{B}$ (with $k_\mathrm{B}$ the Boltzmann constant) this energy is not available, and the macroscopic quantum charge state $Q$ is effectively frozen \cite{Slobodeniuk2013,Idrisov2018} (although not quantized in units of $e$ as long as one channel is perfectly connected \cite{Matveev1995,Nazarov1999,Jezouin2016}).
Consequently, correlations develop between absorbed and emitted electrons.
These culminate if only one transport channel is connected to the island, in which case theory predicts that the electrons entering it and those exiting it are in identical quantum states \cite{Clerk2001,Idrisov2018}.
Effectively, the electronic states within the connected quantum channel are decoupled from the many quasiparticles within the island, despite the fact that the incoming (outgoing) physical electron particles penetrate into (originate from) the island.
Another consequence is that heat evacuation from the island's internal states along the channel is fully suppressed \cite{Slobodeniuk2013}.
In contrast, in the presence of two or more open channels the coherence is lost \cite{Idrisov2018}, and heat evacuation is restored in agreement with the recently observed systematic heat Coulomb blockade of one ballistic channel \cite{Sivre2018}.
Interestingly, the `electron teleportation' proposed in \cite{Fu2010} also relies on the `all-important' Coulomb charging energy of a small island, although combined in that case with Majorana bound states in an altogether different mechanism. 

\begin{figure}
\centering\includegraphics[width=87mm]{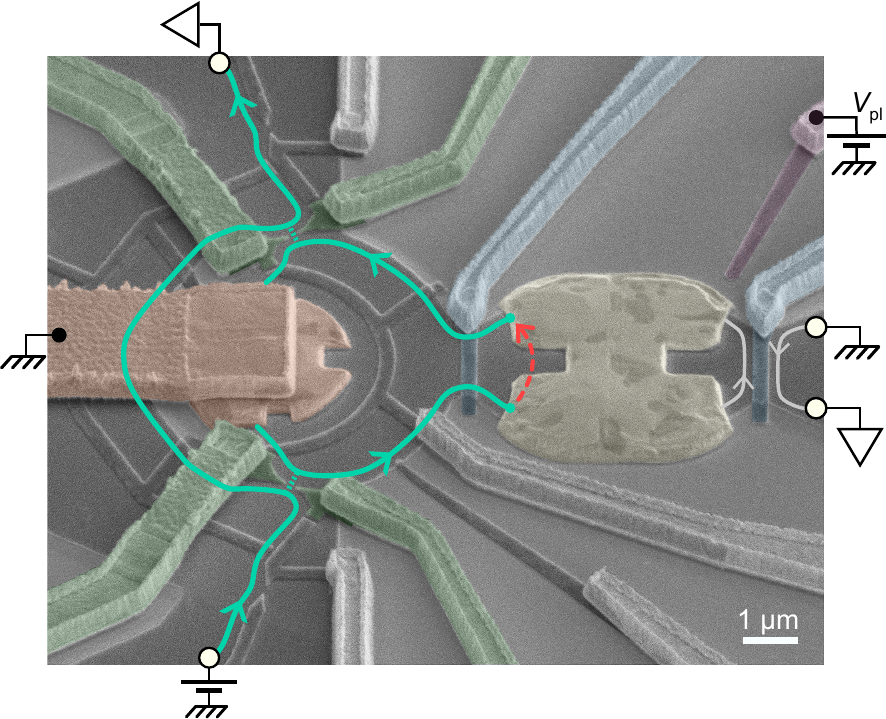}
\caption{
Device e-beam micrograph.
Areas with a Ga(Al)As two-dimensional electron gas underneath the surface appear darker.
The applied perpendicular magnetic field $B\simeq5\,$T corresponds to the integer quantum Hall regime at filling factor two.
Capacitively coupled gates colored green and blue control, respectively, the Mach-Zehnder interferometer beam splitters for the outer quantum Hall edge channel (lines with arrow, here corresponding to the schematic in Fig.~2(b)) and the connection to the floating metallic island (yellow) in good ohmic contact with the buried 2D electron gas.
One of the two MZI outputs is the central small ohmic contact (orange) connected to ground through a suspended bridge.
The second one, larger and located further away, is schematically represented by the top white circle.
The MZI phase difference is controlled through $B$ or the plunger gate voltage $V_\mathrm{pl}$.
The red dashed line visually represents the non-local quantum state transfer across the island, between electrons' injection (starting point) and emission (arrow).
}
\label{fig-sample}
\end{figure}

We demonstrate the high-fidelity replication of electron quantum states across a metallic island through quantum interferences.
For this purpose, an injected current is first split along two separate paths that are subsequently recombined, thereby realizing an electronic Mach-Zehnder interferometer (MZI).
In contrast with usual MZI implementations \cite{Ji2003,Roulleau2007,Litvin2007,Bieri2009,Duprez2019a}, one of the paths can controllably be diverted toward a small floating metallic island (see Fig.~1). 
In that case, any two-path quantum interferences involve both the initial electrons (direct left path) and the reemitted ones (interrupted right path, assuming a perfect contact with the island).
Therefore a high interference visibility directly ascertains a high fidelity of the electron state replication.

\begin{figure}
\centering\includegraphics[width=87mm]{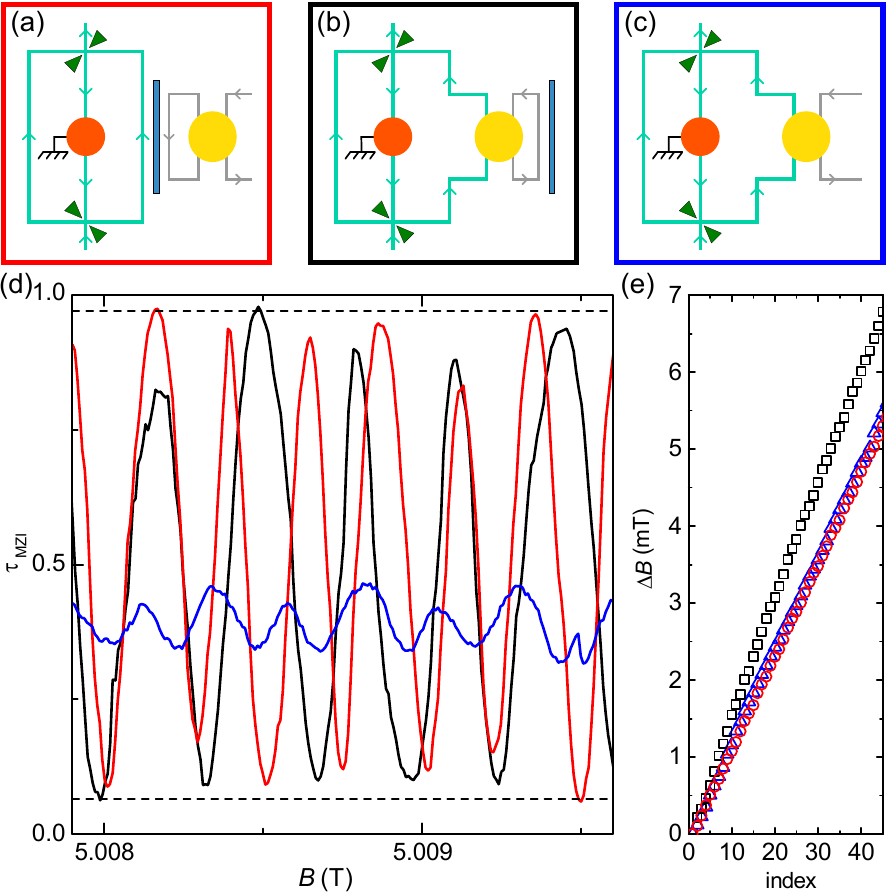}
\caption{
Quantum oscillations versus magnetic field.
(a,b,c), Schematics of implemented MZI configurations.
(d), Fraction $\tau_\mathrm{MZI}$ of the outer edge channel current transmitted across the MZI as a function of $B$.
Continuous lines are measurements performed in the configuration framed by a box of the same color in (a,b,c).
The horizontal black dashed lines represent the $\tau_\mathrm{MZI}$ extrema for the standard and floating island MZI configurations (schematics in (a) and (b), respectively), corresponding to a high quantum oscillations visibility of $\mathcal{V}\sim90\%$.
With a second channel connected to the floating island (configuration shown Fig.~2(c)), the quantum oscillations are strongly reduced to a visibility $\mathcal{V}\sim20\%$, consistent with the separately characterized small residual reflection of $\sim3\%$ (see text and \cite{sm}), and the average $\left\langle\tau_\mathrm{MZI}\right\rangle$ is diminished as part of the current is transmitted across the island toward a remote electrical ground.
(e), Symbols display the magnetic field position of consecutive extrema (both peaks and dips increment the index number).
The larger slope for the floating island MZI configuration (black squares) corroborates the electron quantum state transfer between different injection and emission locations across the floating metallic island.
}
\label{fig-interference}
\end{figure}

A colorized e-beam micrograph of the measured device is shown in Fig.~1.
The sample was nanofabricated from a high-mobility Ga(Al)As two dimensional electron gas, and immersed in a perpendicular magnetic field $B\simeq5\,$T corresponding to the integer quantum Hall filling factor $\nu=2$.
In this regime, two quantum Hall channels co-propagate along the edges (the electron gas was etched away in the brighter areas), and the MZI is formed using only the outer edge channel.
The followed paths are represented by thick lines with arrows for the configuration where one MZI arm goes through the floating metallic island (corresponding schematic shown in Fig.~2(b)).
The two MZI beam splitters, each tuned to half transmission, are realized with quantum point contacts formed by field effect using split gates (colored green; the inner quantum Hall channel, not shown, is fully reflected).
One of the two MZI outputs is the small central metallic electrode (orange), which is grounded through a suspended bridge.
The quantum interferences are characterized by the oscillations of the current transmitted to the second MZI output formed by a much larger electrode 60\,$\mu$m away (represented in Fig.~1 by the top white circle), while sweeping either the magnetic field $B$ or the voltage $V_\mathrm{pl}$ applied to a lateral plunger gate (purple).
The floating metallic island (yellow) consists of $2\,\mu\mathrm{m}^3$ of a gold-germanium-nickel alloy diffused into the Ga(Al)As heterojunction by thermal annealing.
From the typical metallic density of states of such metals $\nu_\mathrm{F}\approx10^{47}\,\mathrm{J}^{-1}\mathrm{m}^{-3}$ ($1.14\times10^{47}$ for gold, the main constituent), the electronic dwell time is $\tau_\mathrm{D}\approx60\,\mu$s.
This is much longer, by more than three orders of magnitude, than the energy relaxation and phase decoherence times of electrons observed in similar metals, which is at most in the $20$\,ns range \cite{Pierre2003,Bauerle2005}.
In the absence of Coulomb-induced correlations, no interferences would therefore be expected from the reemitted electrons, by a wide margin. 
The gates barring the broad way on each side of the floating island (blue) are normally tuned to either fully reflect or fully transmit the outer edge channel, in order to implement the MZI configurations schematically represented Figs.~2(a,b,c).
Note that the second (inner) quantum Hall edge channel is always completely reflected at the barring gate, and can therefore be ignored \cite{Idrisov2018}.
The island charging energy $E_\mathrm{C}\simeq k_\mathrm{B}\times0.3\,$K was obtained from standard Coulomb diamond measurements (in a specifically tuned tunnel regime, see Fig.~3(b) and \cite{sm}).
At the experimental electronic temperature $T\simeq10\,$mK (measured on-chip from shot noise \cite{Iftikhar2016}), the criterion $k_\mathrm{B}T\ll E_\mathrm{C}$ for fully developed Coulomb-induced correlations is therefore well verified.
Note the previous experiments performed in the opposite `high-temperature' regime $k_\mathrm{B}T\gg E_\mathrm{C}$ of negligible Coulomb correlations, in which case, unsurprisingly, a complete quantum decoherence \cite{Roulleau2009} and energy relaxation \cite{Altimiras2010} of electrons were observed with a single connected channel.
Finally, the transparency of the contact between the floating island and the outer quantum Hall edge channel plays an essential role since, if it is poor, many electrons would simply be reflected at the interface.
Here, $\gtrsim97$\% of the incoming current penetrates into the floating island \cite{sm}, which is also ascertained by the striking changes of behavior detailed later.

In Fig.~2, we show illustrative MZI oscillations versus $B$ of $\tau_\mathrm{MZI}$, the fraction of outer edge channel current transmitted across the device.
The measurements were performed in the three configurations depicted in Figs.~2(a,b,c).
The red continuous line in Fig.~2(d) corresponds to a standard electronic MZI, with the floating metallic island bypassed (schematic in Fig.~2(a)).
In that case, the oscillations are of high visibility $\mathcal{V}\equiv\left(\tau_\mathrm{MZI}^\mathrm{max}-\tau_\mathrm{MZI}^\mathrm{min}\right)/\left(\tau_\mathrm{MZI}^\mathrm{max}+\tau_\mathrm{MZI}^\mathrm{min}\right)\approx90\%$ and, as expected for the Aharonov-Bohm phase, the magnetic field period of $241\pm3\,\mu$T (red symbols in Fig.~2(e) show consecutive extrema positions) closely corresponds to one flux quantum ($241\,\mu\mathrm{T}\times S\simeq0.98h/e$ using the nominal area $S\simeq16.8\,\mu\mathrm{m}^2$).
A small asymmetry in the $\tau_\mathrm{MZI}$ data (the average is slightly above $0.5$) results from a small reflection of the outer edge channel on the grounded central ohmic contact (of $\approx5\%$, see \cite{sm}).
The black continuous line in Fig.~2(d) was measured with the right MZI arm deviated to go through the floating ohmic island (edge channel paths displayed in Fig.~1, and schematic in Fig.~2(b)).
We observe first that the quantum interferences' visibility remains of the same high amplitude, which corresponds to a perfect fidelity (at experimental accuracy) of the replicated quantum states imprinted on the electrons reemitted from the island, in agreement with low temperature predictions \cite{Clerk2001,Idrisov2018}.
Second, the magnetic field period of $305\pm4\,\mu$T is found to be larger than in the standard MZI configuration of Fig.~2(a) (see black symbols in Fig.~2(e)).
This increase is \textit{opposite} to the reduction that would be expected from the Aharonov-Bohm period with the larger surface enclosed by the outer channel path and the inner boundary of the floating metallic island (see \cite{sm} for a graphical representation, $S\simeq18.4\,\mu\mathrm{m}^2$ would correspond to an Aharonov-Bohm period of $225\,\mu\mathrm{T}\simeq h/eS$).
Such opposite evolution and relatively important discrepancy (36\%) establish that the MZI phase does not reduce to the usual Aharonov-Bohm phase acquired by a single electron propagating along two different paths.
Instead, the larger period corroborates the transfer of the electrons' state across the island, thereby amputating the electron path from a section (the 2DEG/metal interface) and making the Aharonov-Bohm notion of enclosed surface ill-defined.

\begin{figure}[htb]
\centering\includegraphics[width=87mm]{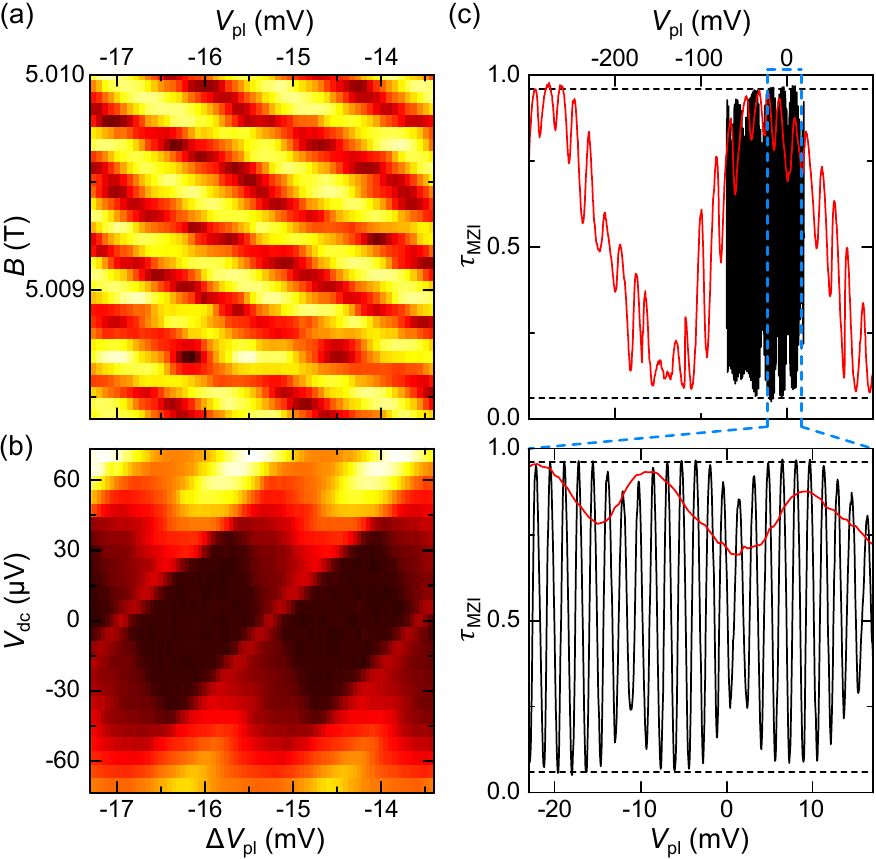}
\caption{
Quantum phase versus island charge. 
(a), Color plot of $\tau_\mathrm{MZI}(B,V_\mathrm{pl})$ in the floating island MZI configuration (schematic in Fig.~2(b)), with the larger values shown brighter, which establishes the equivalent role of $B$ and $V_\mathrm{pl}$.
(b), Coulomb diamonds characterization of the floating island (larger differential conductance shown brighter, with the island here weakly coupled on both sides and $V_\mathrm{dc}$ the applied dc bias voltage).
A comparison with panel (a), plotted using the same $V_\mathrm{pl}$ scale, reveals that the addition of a charge of $e$ on the island precisely corresponds, in the floating island MZI configuration, to an electron quantum phase of $2\pi$ (one quantum oscillation period).
(c), The top and bottom panels display measurements of $\tau_\mathrm{MZI}(V_\mathrm{pl})$ with the device set in the floating island MZI configuration (black line) and in the standard MZI configuration (red line, schematic in Fig.~2(a)).
The MZI oscillations' period in $V_\mathrm{pl}$ is shorter by a factor of $1/160$ when the island is connected.
Note an additional modulation of fixed period ($\approx15$\,mV).
}
\label{fig-nonEq}
\end{figure}

The blue continuous line in Fig.~2(d) was measured with one MZI arm going through the floating island, and in the presence of a second electronic channel connected to it (configuration schematically displayed in Fig.~2(c)).
We find strongly suppressed conductance oscillations corresponding to a full decoherence of the electrons going through the island.
The residual visibility $\mathcal{V}\lesssim0.2$ is consistent with the proportion $1-\tau_\mathrm{island}\lesssim3\%$ of reflected electrons, not penetrating into the island.
Indeed, the MZI contribution of the reflected electrons at small $1-\tau_\mathrm{island}\ll1$ reads $\mathcal{V}_0(4/3)\sqrt{1-\tau_\mathrm{island}}\lesssim0.21$, with $\mathcal{V}_0\approx90\%$ the MZI visibility in the standard configuration \cite{Roulleau2009,sm}.
The magnetic field period of $246\pm4\,\mu$T for these smaller oscillations (see blue symbols in Fig.~2(e)) is found close to the period observed in the standard MZI configuration shown in Fig.~2(a), suggesting that the residual reflections take place at the level of the barring gate (colored blue, left of island in Fig.~1).
Note that the average $\langle\tau_\mathrm{MZI}\rangle\simeq0.39$ is shifted below $0.5$ because part of the injected current is evacuated toward a remote electrical ground through the second channel connected to the floating island ($\langle\tau_\mathrm{MZI}\rangle=0.375$ expected from current conservation for a floating island and a central ohmic contact both perfectly connected).

We now investigate the relation between the island's charge and the electron phase shift associated with the quantum state transfer.
For this purpose, Fig.~3 focuses on the influence on $\tau_\mathrm{MZI}$ of the voltage $V_\mathrm{pl}$ applied to a plunger gate (colored purple in Fig.~1) which is relatively far from the MZI outer quantum Hall channel, but close to the island.
The equivalent role on the MZI phase of $V_\mathrm{pl}$ and $B$ is first directly established, in Fig.~3(a), with the device set in the floating island MZI configuration (schematic in Fig.~2(b)).
Figure~3(b) displays Coulomb diamond measurements of the conductance across the island as a function of the same plunger gate voltage $V_\mathrm{pl}$, with here the island weakly connected through tunnel barriers such that $Q$ is quantized in units of $e$ (only in that specific case) and without two-path interferences (see device schematic in \cite{sm}).
Remarkably, the MZI gate voltage period in Fig.~3(a) precisely matches the Coulomb diamonds' period in Fig.~3(b), as can be seen by directly comparing the two panels plotted using the same $V_\mathrm{pl}$ scale.
In the floating MZI limit of strongly connected channels $Q=eV_\mathrm{pl}/\Delta$, with $\Delta\simeq1.7\,$mV the Coulomb diamond period \cite{Matveev1995,Nazarov1999,Jezouin2016}.
A quantum phase shift of $2\pi Q/e$ therefore applies to the transferred electrons, as specifically predicted theoretically \cite{Clerk2001,Idrisov2018}, and in agreement with Friedel's sum rule.
Comparing with the device set in the standard MZI configuration, we show in Fig.~3(c) that the $\tau_\mathrm{MZI}$ oscillations (red line) are of identical maximum visibility $\mathcal{V}\simeq90\%$ than with one arm going through the metallic island (black line), as also seen versus magnetic field in Fig.~2(d).
However, the $V_\mathrm{pl}$ period is increased by a large factor of 160, from $1.7\,$mV to 270\,mV, which reflects the weak coupling of the plunger gate voltage to the MZI outer edge channel (see \cite{sm} for an extended $V_\mathrm{pl}$ range).
This provides a final evidence that the electrons contributing to the quantum oscillations in the floating island configuration indeed penetrate into the metal.
Note the presence of an additional, smaller signal of fixed period 15\,mV visible in both configurations (in the form of direct oscillations or of an amplitude modulation), which might originate from the progressive charging of a nearby defect.

This experimental work demonstrates that the Coulomb interaction has two facets.
It can both destroy and preserve quantum effects.
Although a metallic island is often pictured as a floating reservoir of uncorrelated electrons \cite{Buttiker1988a,deJong1996}, we establish that a high-fidelity electron quantum state transfer can take place across it, enforced by the Coulomb charging energy.
This provides a mean to overcome limitations imposed by the decoherence of individual electrons.
Moreover, the observed universal $2\pi$ electron phase shift for one elementary charge $e$ on the island allows for a strong entanglement of single-electron states, both between themselves or with other quantum degrees of freedom, with a negligible loss of coherence.
Such controllable, strong-coupling mechanism constitutes a key element in the context of quantum Hall edges envisioned as platforms for the manipulation and transfer of quantum information via propagating electrons \cite{Bertoni2000,Ionicioiu2001,Stace2004,Bocquillon2014,Glattli2017,Bauerle2018,Duprez2019a}.
In particular, it is remarkably well suited to implement quantum gates for these `flying qubits', such as the CNOT proposal involving a conditional phase shift of $\pi$ described in \cite{Glattli2017}.

\subsection*{Acknowledgments}
This work was supported by the French RENATECH network, the national French program `Investissements d'Avenir' (Labex NanoSaclay, ANR-10-LABX-0035) and the French National Research Agency (project QuTherm, ANR-16-CE30-0010).

E.S. and H.D. performed the experiment and analyzed the data with inputs from A.Aa., A.An. and F.P.;
F.P. fabricated the sample with inputs from E.S and H.D.;
A.C., and U.G. grew the 2DEG;
F.P. led the project and wrote the manuscript with inputs from A.Aa., A.An., E.S., H.D. and U.G.

We thank P.~Brouwer, L.~Glazman, C.~Mora, Y.~Oreg and E.~Sukhorukov for illuminating discussions.

\end{document}


\title{Supplementary Materials for \\`Transferring the quantum state of electrons across a Fermi sea with Coulomb interaction'}

\author{H. Duprez}
\email{These authors contributed equally to this work.}
\affiliation{Centre de Nanosciences et de Nanotechnologies (C2N), CNRS, Univ Paris Sud, Universit\'e Paris-Saclay, 91120 Palaiseau, France}
\author{E. Sivre}
\email{These authors contributed equally to this work.}
\affiliation{Centre de Nanosciences et de Nanotechnologies (C2N), CNRS, Univ Paris Sud, Universit\'e Paris-Saclay, 91120 Palaiseau, France}
\author{A. Anthore}
\affiliation{Centre de Nanosciences et de Nanotechnologies (C2N), CNRS, Univ Paris Sud, Universit\'e Paris-Saclay, 91120 Palaiseau, France}
\affiliation{Univ Paris Diderot, Sorbonne Paris Cit\'e, 75013 Paris, France}
\author{A. Aassime}
\affiliation{Centre de Nanosciences et de Nanotechnologies (C2N), CNRS, Univ Paris Sud, Universit\'e Paris-Saclay, 91120 Palaiseau, France}
\author{A. Cavanna}
\affiliation{Centre de Nanosciences et de Nanotechnologies (C2N), CNRS, Univ Paris Sud, Universit\'e Paris-Saclay, 91120 Palaiseau, France}
\author{U. Gennser}
\affiliation{Centre de Nanosciences et de Nanotechnologies (C2N), CNRS, Univ Paris Sud, Universit\'e Paris-Saclay, 91120 Palaiseau, France}
\author{F. Pierre\thanks{frederic.pierre@c2n.upsaclay.fr}}
\email[e-mail: ]{frederic.pierre@c2n.upsaclay.fr}
\affiliation{Centre de Nanosciences et de Nanotechnologies (C2N), CNRS, Univ Paris Sud, Universit\'e Paris-Saclay, 91120 Palaiseau, France}

\maketitle

\section*{Materials and Methods}

\subsection{Sample and measurement setup}

{\noindent\textbf{Sample.}}
The device is patterned by standard e-beam lithography, dry etching and metallic deposition on a two-dimensional electron gas (2DEG) located 95\,nm below the surface of a Ga(Al)As heterojunction.
The 2DEG is of density $2.5\times10^{11}\,\mathrm{cm}^{-2}$ and mobility $2.5\times10^6\,\mathrm{cm}^2\mathrm{V}^{-1}\mathrm{s}^{-1}$, separately characterized on a different chip.
The micron-scale metallic islands (colored orange and yellow in Fig.~1) are made of nickel (30\,nm), gold (120\,nm) and germanium (60\,nm).
A good ohmic contact was established with the buried 2DEG by thermal annealing at 440\,$^\circ$C for 50\,s.
The left and right MZI arms were designed to be symmetric in the standard MZI configuration schematically shown in Fig.~2(a), both of length $L\approx7.3\,\mu$m.
The quantum Hall edge path along the right MZI arm was designed such that its length remains the same whether the floating island is connected (as schematically illustrated in Fig.~2(b), including the inserted metallic island as illustrated with the green continuous line in Fig.~1) or not connected (as schematically illustrated in Fig.~2(a)).
In the standard MZI configuration, the surface enclosed between the two MZI arms is $S\simeq16.8\,\mu\mathrm{m}^2$ (area colored red in Supplementary Fig.~S1(a)).
In the floating island MZI configuration, the surface delimited by the two MZI edge path and the floating metallic island is $S\simeq18.4\,\mu\mathrm{m}^2$ (area colored red in Supplementary Fig.~S1(b), 10\% larger than in the standard MZI configuration).\\

\begin{figure*}
\centering\includegraphics[width=116mm]{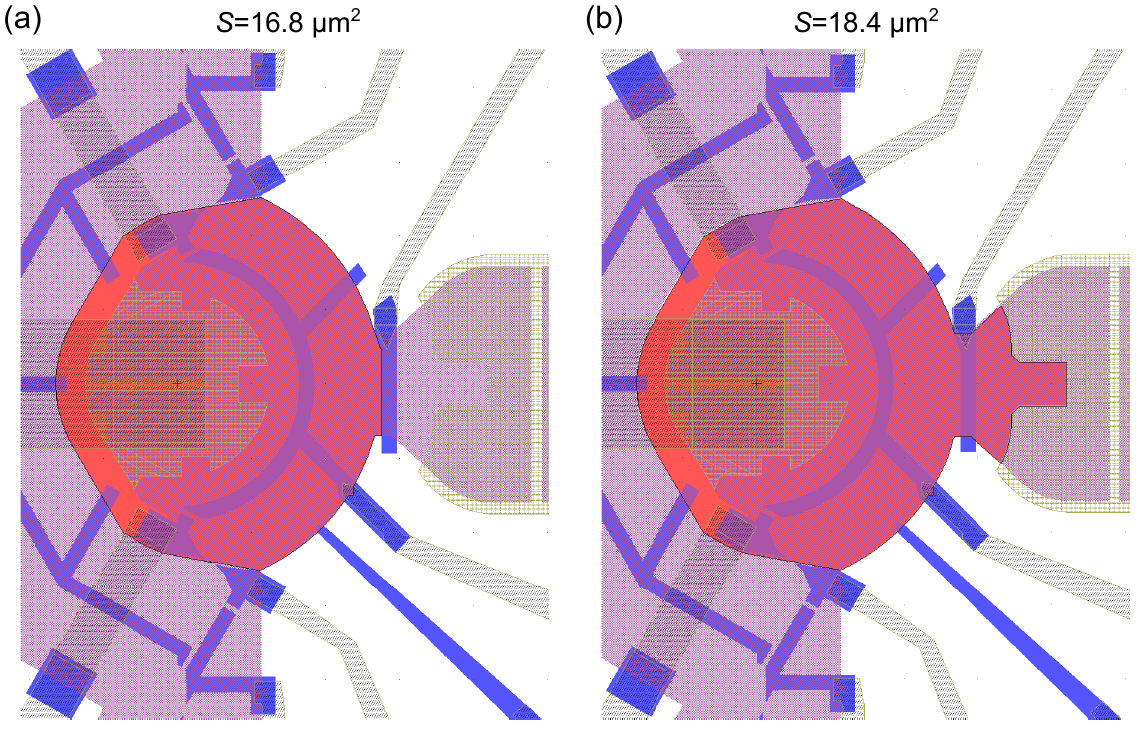}
\caption{
MZI enclosed surface.
The considered areas for the MZI enclosed surface $S$ quoted in the manuscript are highlighted in red for the standard MZI configuration (a) and for the floating island MZI configuration (b). 
}
\label{SIFig-areas}
\end{figure*}

{\noindent\textbf{Experimental setup.}} 
The device is cooled down to 10\,mK in a cryofree dilution refrigerator and connected by electrical lines including several filtering and thermalization stages.
The electron temperature is obtained from quantum shot-noise measurements across a quantum point contact set to half transmission for the outer edge channel (the top MZI beam splitter in Fig.~1, with the bottom MZI beam splitter and the uncolorized left gate set, respectively, to fully transmit and reflect the outer edge channel).
See [22] for further details on this experimental setup.\\

\subsection{Device characterization}

{\noindent\textbf{Interfaces between 2DEG and ohmic contacts.}}
The grounded central ohmic contact's quality is characterized by the ratio of reflected over impinging current. 
Ideally, there should be no reflected current (with a perfect metal/2DEG interface and a zero resistance path to electrical ground). 
In practice, if the impinging current is carried only by the outer edge channel (used for the interferometer), the reflected current is always found below 8\%.
Typically the outer edge channel reflection is $\sim5$\%, with a noticeable dependence on the voltage applied to nearby gates.
Note that the inner quantum Hall edge channel (not used for the MZI) is almost fully reflected (96\%) from the grounded central ohmic contact.
A similar characterization can be performed on the floating island with both barring gates (blue in Fig.~1) set to fully transmit the outer edge channel.
Ideally, if the two metal/2DEG interfaces are perfect, the exact same amount of outer edge channel current should be detected on the right side of the floating island and on the top MZI electrode (with both MZI beam splitters set to fully transmit the outer edge channel).
We find here that the ratio between these two measured currents deviates from this ideal case by $\sim3\%$, which provides a maximum value for the reflection probability at the metal/2DEG interface on the MZI (left) side of the floating island.\\

{\noindent\textbf{Charging energy.}}
The floating island's charging energy $E_\mathrm{C} = e^2/2C \approx k_B \times 0.3~$K is given by the half-height in drain-source dc bias voltage $V_\mathrm{dc}$ of the measured Coulomb diamonds shown in Fig.~3(b).
Note that for this measurement the two MZI beam splitters were adjusted to fully transmit the outer edge channel, as schematically illustrated in Supplementary Fig.~S2.\\

\begin{figure}
\centering\includegraphics[width=58mm]{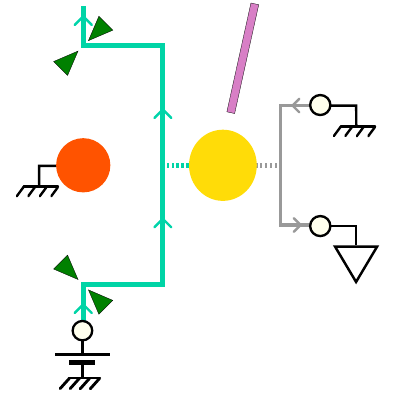}
\caption{
Device configuration for Coulomb diamond measurements. 
}
\label{SIFig-schem_diam}
\end{figure}

\subsection{Supplementary data}

\begin{figure}
\centering\includegraphics[width=87mm]{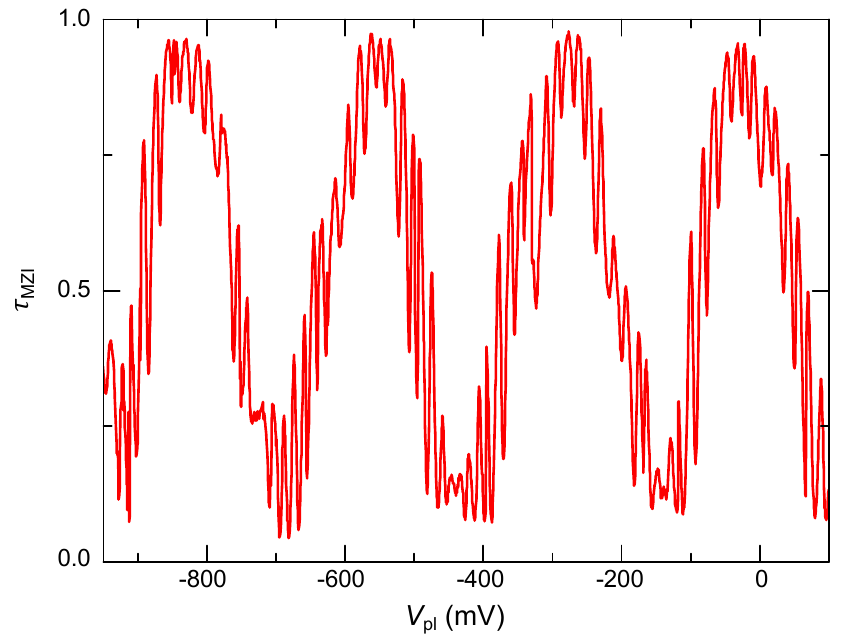}
\caption{
Extended quantum oscillations in standard MZI configuration.
The red continuous line displays the transmitted fraction $\tau_\mathrm{MZI}(V_\mathrm{pl})$ measured in the standard MZI configuration (schematically represented in Fig.~2(a)).
It is the same data displayed in the top panel of Fig.~3(c), but here shown on a larger $V_\mathrm{pl}$ range unambiguously demonstrating their periodic character.
}
\label{SIFig-tauMZI(Vpl)_extended}
\end{figure}

{\noindent\textbf{Quantum oscillations versus plunger gates.}}
The $\tau_\mathrm{MZI}(V_\mathrm{pl})$ oscillations in the floating island MZI configuration (schematic in Fig.~2(b)), previously shown in Fig.~3(c), are displayed in Supplementary Fig.~S3 over a broader $V_\mathrm{pl}$ range of 1\,V, including 4 periods in order to clearly establish their periodic character.\\

\begin{figure}
\centering\includegraphics[width=87mm]{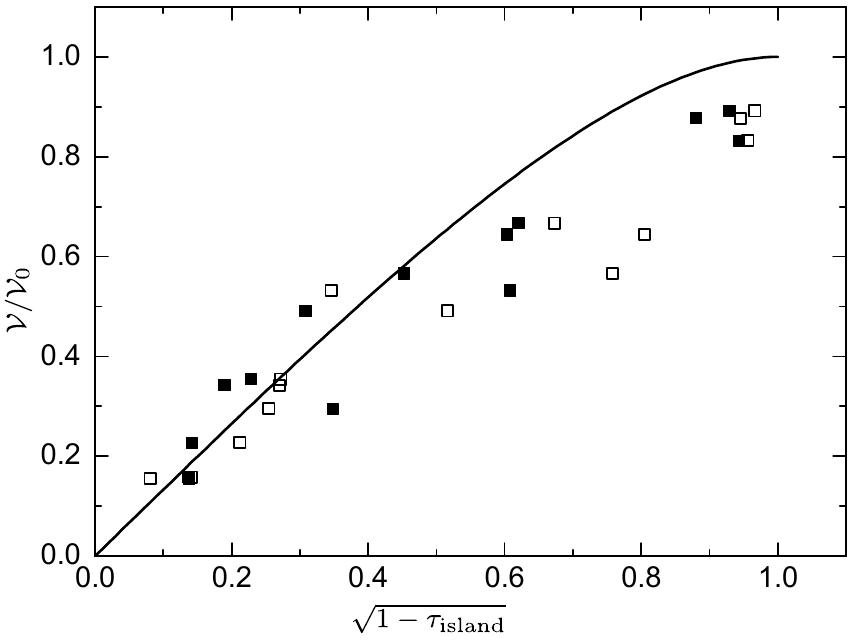}
\caption{
Quantum oscillations versus island coupling in the presence of a second channel.
The continuous line is calculated from Supplementary Eq.~S1.
Symbols represent oscillation visibility measurements performed in the presence of a second channel connected to the island, as a function of the transmission probability $\tau_\mathrm{island}$ of the MZI outer edge channel toward the metallic island.
Whereas $\tau_\mathrm{island}$ is very close to one in the configuration schematically shown in Fig.~2(c) and discussed in the manuscript (only limited by the metal/2DEG interface's quality), here its value was controllably reduced by tuning the voltage applied to the left blue barring gate (see Fig.~1).
Two different methods were used to extract $\tau_\mathrm{island}$, displayed with open and full symbols (see text).
The used value for the standard MZI visibility $\mathcal{V}_0\simeq89\%$ was specifically determined at $\tau_\mathrm{island}=0$ under the same experimental conditions at the end of this sequence of measurements. 
}
\label{SIFig-sqrt(1-tau)}
\end{figure}

{\noindent\textbf{Visibility versus transmission probability.}}
Here we test the dependence of the MZI oscillations' visibility $\mathcal{V}$ with the transmission probability $\tau_\mathrm{island}$ toward the metallic island, in the presence of a second channel well connected to a grounded electrode.
This configuration corresponds to the schematic shown Fig.~2(c) but with a controllable transmission $\tau_\mathrm{island}$ of the inner blue barring gate (left in Fig.~1).
Assuming that there is no coherence between injected and emitted electrons, as expected independently of $k_\mathrm{B}T/E_\mathrm{C}$ with a second channel [9], one finds that the visibility is reduced by the factor:
\begin{equation}
\mathcal{V}/\mathcal{V}_0=\frac{2\sqrt{1-\tau_\mathrm{island}}(1+\tau_\mathrm{island})}{2+\tau_\mathrm{island}},
\end{equation}
with respect to the visibility $\mathcal{V}_0$ in the standard MZI configuration ($\tau_\mathrm{island}=0$). 
For a small reflection probability ($1-\tau_\mathrm{island}\ll1$), this expression reduces to $(4/3)\sqrt{1-\tau_\mathrm{island}}$ as pointed out in the manuscript.
Note the difference with the previously established relation $\mathcal{V}/\mathcal{V}_0=\sqrt{1-\tau_\mathrm{island}}$ for a single connected channel in the high temperature regime $k_\mathrm{B}T\gg E_\mathrm{C}$ [23].

Experimentally, the comparison is here limited by the difficulty to reliably adjust $\tau_\mathrm{island}$ with a continuous barring gate (instead of a split gate normally used to form a QPC), due to rapid variations with the applied gate voltage.
Supplementary Fig.~S4 shows as open and closed symbols the result of two different approaches to extract $\tau_\mathrm{island}$, with the continuous line representing the prediction of Supplementary Eq.~S1.
In the first approach (full symbols), we determine $\tau_\mathrm{island}$ from the average transmitted current across the island and toward the MZI output.
The drawback in that case is that the obtained values are influenced by the precise amount of reflection on the grounded central ohmic contact, as well as on the precision of our adjustment of the QPC beam splitters.
In the second approach (open symbols), we determine $\tau_\mathrm{island}$ versus barring gate voltage $V_\mathrm{g}$ with the two beam splitter QPCs set to fully transmit the outer edge channel (following the procedure described above to determine the interface quality).
Then we assume that it is possible to correct for the capacitive crosstalk when changing back the beam splitters transmission to one half by a fixed offset in $V_\mathrm{g}$.
The latter is then simply obtained by adjusting the recognizable pattern of peaks and dips locations in $\tau_\mathrm{island}(V_\mathrm{g})$.
As seen in Supplementary Fig.~S4, we find a reasonable agreement given experimental uncertainties, which further establishes our understanding of the $\lesssim0.2$ visibility observed in Fig.~2(d) (blue line) when the device is set in the MZI configuration shown in Fig.~2(c).